# Spatially Selective and Reversible Doping Control in Cuprate Films

Seongshik Oh[a)], Joseph A. Bonetti[b)], Kevin Inderhees, D. J. Van Harlingen and J. N. Eckstein

Department of Physics, University of Illinois, Urbana, IL 61801

[a)] Present address: National Institute of Standards and Technology, Boulder, CO 80305

[b)] Present address: Jet Propulsion Laboratory, California Institute of Technology, Pasadena, CA 91109

We describe a reversible, spatially-controlled doping method for cuprate films. The technique has been used to create superconductor-antiferromagnetic insulator-superconductor (S-AFI-S) junctions and optimally doped superconductor-underdoped superconductor-optimally doped superconductor (OS-US-OS) cuprate structures. We demonstrate how the S-AFI-S structure can be employed to reliably measure the transport properties of the antiferromagnetic insulator region at cryogenic temperatures using the superconductors as seamless electrical leads. We also discuss applied and fundamental issues which may be addressed with the structures created with this doping method. Although it is implemented on a cuprate film ($YBa_2Cu_3O_{7-\delta}$) in this work, the method can also be applied to any mixed-valence transition metal oxide whose physical properties are determined by oxygen content.

Spatially selective doping is one of the key technologies employed in modern semiconductor electronics.[1] The basic principle is to protect selected regions by a diffusion (or implantation) barrier and flood expose the whole sample to a dopant source. The barrier-protected regions remain undoped, while the unprotected regions get populated with the desired dopant.

Another material whose electronic properties are altered by doping is the high critical temperature superconducting cuprates, although their doping mechanism is different from that of semiconductors. In cuprates, the doping level is usually controlled by the level of oxygen content or by cation substitution. However, photodoping[2,3,4,5] also creates carriers and, with focusing optics, can be used to dope specific regions. Currently, photodoping is the most common way to control spatially the doping level of cuprate films. Photodoping, however, is not stable at room temperature and is effective only on a limited group of cuprates (mostly the $YBa_2Cu_3O_{7-\delta}$ family).[2] In this Letter, we present a spatially-controlled doping method which is stable at room tempeature and applicable to any cuprate. Moreover, the technique can be applied to any mixed-valence transition metal oxide whose physical properties change with oxygen content.[6,7,8] Using this technique, we have created in-plane S-AFI-S and OS-US-OS doping profiles and demonstrated that the S-AFI-S structure can be employed to measure the transport properties of the AFI at cryogenic temperatures. The ability to create stable, locally doped regions in a cuprate material opens up many possible microelectronic and device applications. In addition, fundamental questions can be explored. For instance, S-AFI-S cuprate junctions have attracted considerable attention in recent years due to the possibility of a long range proximity coupling.[9] Also, the OS-US-OS structures could be employed as an ideal way to study the pseudogap state of the underdoped regions.[10]

The main idea behind this new technique is to combine a microfabricated oxygen diffusion barrier (for example, Si) with appropriate thermal treatment to oxygenate or deoxygenate a specific part of the cuprate film. Although this doping control method in cuprates is conceptually similar to that in semiconductors, there is one big difference --- it is completely reversible, whereas doping in semiconductors is not. The whole process to define either S-AFI-S or OS-US-OS structures has been repeated over many cycles without any noticeable sample degradation effect. In addition, each doping profile remained stable with only minor resistance drift over the period of investigation (several weeks).

Detailed process steps to define S-AFI-S or OS-US-OS profiles are described in Fig. 1(a); Fig. 1(b) is an alternative procedure that could be used to produce the same final device pattern. Step 1 is depositing the cuprate film. We used 25 nm thick $YBa_2Cu_3O_{7-\delta}$ (Y123) films grown by pulsed laser deposition on $SrTiO_3$(100) substrates. Step 2 is to define the device structure using an appropriate patterning technique. We have selected a bar geometry with a narrow Si-overlayer covering part of the film; a top view is shown in Fig. 2(a). First, the bar pattern in Fig. 2(a) is defined by deep ultraviolet (UV) lithography and wet chemical etching (0.02 % $HNO_3$:$H_2O$) using PMMA (polymethyl-methacrylate) as the resist. Then, e-beam lithography and Si lift-off (50 nm thick) is used to define the narrow Si line with the previous PMMA pattern as the resist. The Si-overlayer behaves as a diffusion barrier against oxygen in later annealing steps. Any non-conducting material which is inert enough to the oxygen environment could be used instead of Si. Step 3 is to anneal the Si-overlayered sample in vacuum (<$10^{-7}$ Torr) to reduce oxygen content over the entire sample. It is critical to maintain a good vacuum, otherwise the sample can degrade noticeably after each annealing. The annealing time and temperature during this step determines the final doping level under the Si-overlayer. For example, a 10 minute anneal at 450 °C converted the sample into an antiferromagnetic insulator, whereas a 5 minute anneal at the same temperature made the sample just underdoped with $T_c = 40$ K. The final step (step 4) is to anneal the sample at a low temperature (250 °C) in an ozone environment (~$10^{-6}$ Torr) for 30 minutes. During the ozone step, oxygen diffusion is suppressed due to the relatively low annealing temperature. Thus, the region underneath the Si-layer remains oxygen deficient while the regions outside the Si-overlayer are converted into near-optimally doped material ($T_c = 80$ K) since the diffusion level is still adequate to re-oxygenate these bare regions. Unless the temperature and the time are accurately optimized, the actual oxygen deficient region would be substantially narrower than the Si-overlayer. This introduces an error in the estimation of electric field for the transport data below. In this work Si-overlayer widths of 5.6 µm, 11.8 µm, 25 µm, 43 µm and 174 µm were used to test the width dependence.

In between each processing step, four point probe measurements were taken to check the junction resistance. Fig. 2 shows the temperature dependence of the junction resistance before and after the thermal process. Before Step 3, junctions, one with $d = 5.6$ µm Si-overlayer and the other without any Si-overlayer ($d = 0$), show superconducting transitions at around 80 K (see Fig. 2(b)). However, after they are thermally treated (Step 4) and presumably transformed into S-AFI-S junctions, fascinating differences show up between the Si-overlayered junction and the control junction without any Si-overlayer (see Fig. 2(c)). While the $d = 0$ junction still behaves superconducting below 80 K, $d = 5.6$ µm junction has become insulating. In order to make sure that this insulating behavior is not due to sample damage but due to oxygen deficiency, the same sample was later exposed to ozone ($10^{-6}$ Torr) at a high temperature (450 °C) for 30 minutes, and all the junctions became superconducting again at the same critical temperature (80 K) as the original film (see Fig. 2(d)). This verifies that the earlier insulating behavior was due to oxygen deficiency

S-AFI-S structures provide an ideal way to electrically characterize the AFI region. In general, transport measurement of a highly insulating antiferromagnetic cuprate is difficult to perform at cryogenic temperatures due to poor Ohmic contact to the insulator. The S-AFI-S structure eliminates this problem. Once the sample is cooled below the critical temperature of the outer superconductors, the superconducting parts can be used as resistanceless contact leads. Contact resistance between external wiring and the superconducting leads is negligible because the contact is between two metallic materials. The contact between superconducting leads and the antiferromagnetic insulator is just from doping difference and so there is no parasitic contact resistance at the junction. Since the resistance across the AFI region is very large (higher than a Giga Ohm) at low temperatures, four point measurement is not reliable and voltage biased two point measurement should be used instead. Although two point measurements usually suffer from the contact resistance problem, the S-AFI-S system is nearly free from such an issue once the small contact resistance between external leads and superconducting pads is subtracted.

Fig. 3 shows current vs. electric field ($I$-$E$) measured at 4.2 K across four different S-AFI-S junctions, where the electric field is defined as the voltage ($V$) across the Si-overlayers divided by their widths. Since the actual oxygen deficient line width is smaller than that of the Si-overlayer due to non-zero oxygen diffusion length, the actual electric fields must be larger than the values defined this way. This is the main reason why curves in Fig. 3 do not overlap. Apart from that, Fig. 3 demonstrates a severe nonlinearity between the current and the electric field, especially at high electric fields. At high fields, it gives a power-law dependence of $I \sim E^{4.8}$. The physical origin of this specific power-law relationship is not known, and a more in-depth study will be required to get a better understanding. Still this measurement demonstrates how the seamless electrical contact method can be employed to obtain reliable transport data for the AFI state of the cuprates.

As we mentioned above, if the sample is annealed for less time in Step 3, the Si-overlayered region can be made into an underdoped superconductor, forming an OS-US-OS doping profile. Fig. 4 shows the temperature dependence of resistance for three OS-US-OS junctions with different Si-overlayer widths. The largest width junction ($d = 174$ μm) with Si-overlayer wide enough to cover almost the entire region between the voltage leads, shows only one transition at around 40 K, whereas other junctions with narrower Si-overlayers show two transitions, one near 80 K and another near 40 K. This implies that the material beneath the Si has become underdoped with $T_c$ of 40 K while the remaining material is near optimally doped with $T_c$ of 80 K. At temperatures below $T_c$ of outer conductors, only the Si-overlayered region of each junction contributes to the measured resistance.

In conclusion, we have devised a spatially controlled doping method for cuprate materials which is reversible and stable at room temperature. Using this method, we have demonstrated the ability to create S-AFI-S and OS-US-OS structures. These structures provide many exciting avenues for applications and also for fundamental physics research. Finally since the technique can be transferred to any mixed valence transition metal oxide, it has the potential to play a role in important applications in other materials such as chromium oxides[7] and iron oxides.[8]

We acknowledge support from the U. S. Office of Naval Research under grant N00014-00-1-0840 and the U. S. Department of Energy, Basic Energy Sciences, through the Frederick Seitz Materials Research Laboratory award #DEFG02-91ER45439.

**Figure Captions**

FIG. 1. (Color online). (a) Detailed process steps to obtain S-AFI-S or OS-US-OS junction. Step 1: Cuprate film grown on a substrate. Step 2: Si-overlayer deposited using lithography and liftoff; top view of the device after this step is shown in Fig. 2(a). Step 3: High temperature (450 °C) annealing in vacuum to reduce oxygen content in the entire film. Step 4: Low temperature (250 °C) annealing in ozone to oxidize outside Si-overlayered region. (b) An alternative procedure, which we have not used but is equally valid, to make the same final pattern. This second procedure is different from (a) only in Step 3 and 4 in that high temperature annealing to reduce oxygen is done before the overlayer is deposited. This procedure may be useful if the overlayer is not compatible with the high temperature process.

FIG. 2. (Color online). Resistance vs. Temperature plots for S-AFI-S junctions at different stages. (a) Top view of the junction structure and the four point probe geometry. Si-overlayer of width $d$ is shown. (b) Before Step 3 in Fig. 1, both junctions with and without Si-overlayer are superconducting near 80 K. (c) After Step 4, Si-overlayered region ($d$ = 5.6 µm) has become insulating, while the control junction without any Si-overlayer ($d$ = 0) remains superconducting with almost the same $T_c$ as in (b). (d) After the sample is exposed to ozone at a high temperature, the insulating sample has become superconducting again with $T_c$ of 80 K. This verifies that the insulating behavior in (c) is due to oxygen deficiency and not due to sample damage.

FIG. 3. Current vs. Electric field plot for S-AFI-S junctions. Note the severe nonlinearity. Straight lines at large electric field correspond to a power law dependence of $I \sim E^{4.8}$. Electric field is defined as the voltage divided by the width of each Si-overlayer. Curves of different widths do not overlap. This implies that the actual oxygen deficient line widths are smaller than the overlayer widths due to non-zero oxygen diffusion length. This effect can be minimized by optimizing the oxidation temperature and time during Step 4 in Fig. 1(a).

FIG. 4. (Color online). Resistance vs. Temperature for OS-US-OS junctions; (a) and (b) differ only in the vertical axis format. For $d$ = 25 μm and 5.6 μm, the transitions at the higher temperature (80 K) are from the outer conductors (OS) and the transitions at the lower temperature (40 K) are from the central region under the Si-overlayer (US). For $d$ = 174 μm, only one transition occurs at the lower temperature because the resistance is dominated by the wide underdoped superconductor (US).

**Fig. 1**

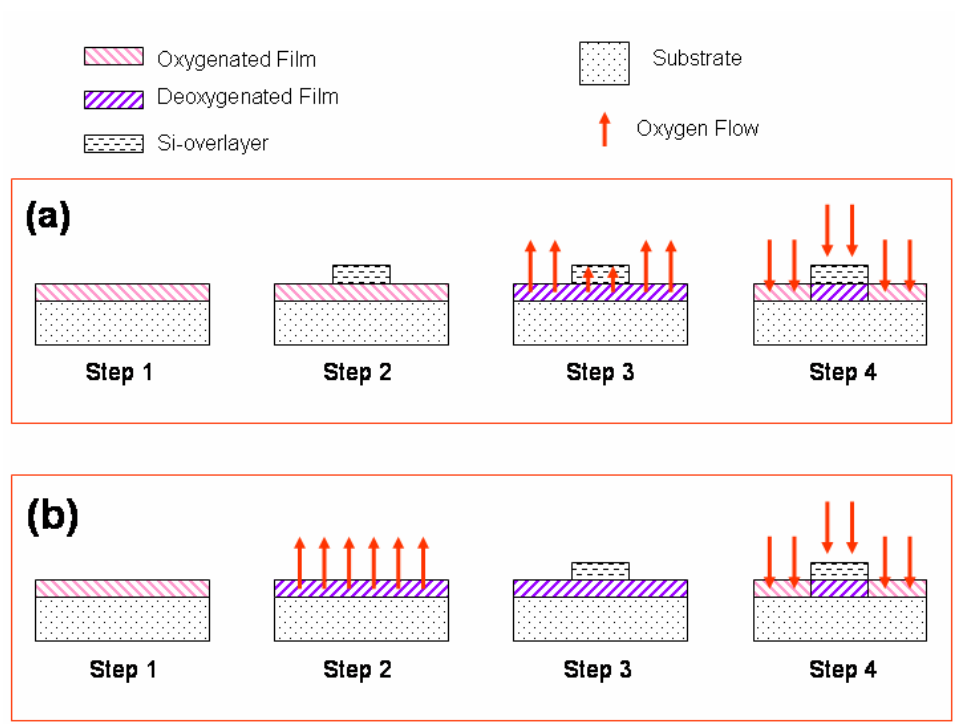

**Fig. 2**

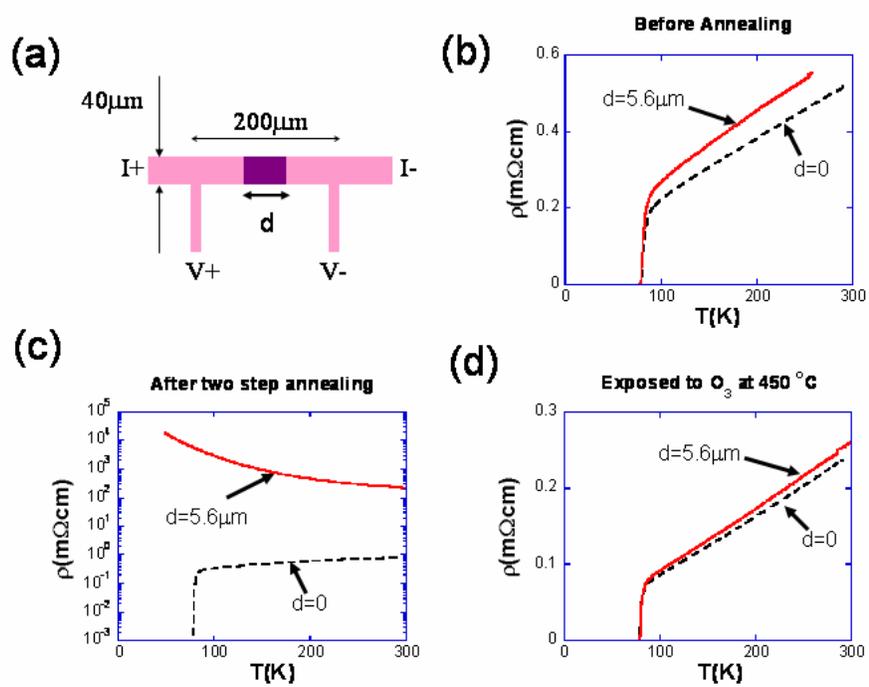

**Fig. 3**

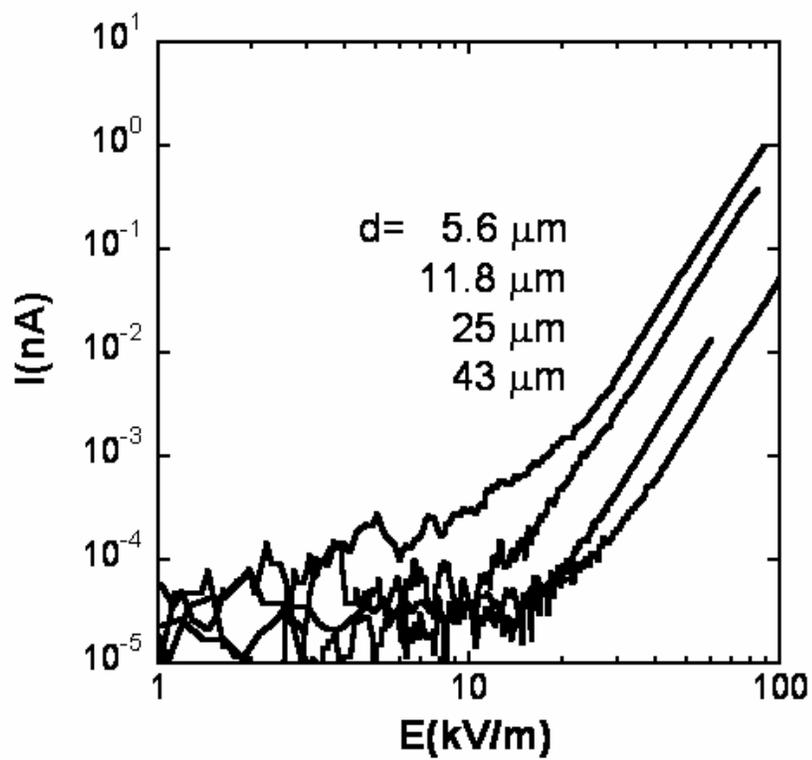

**Fig. 4**

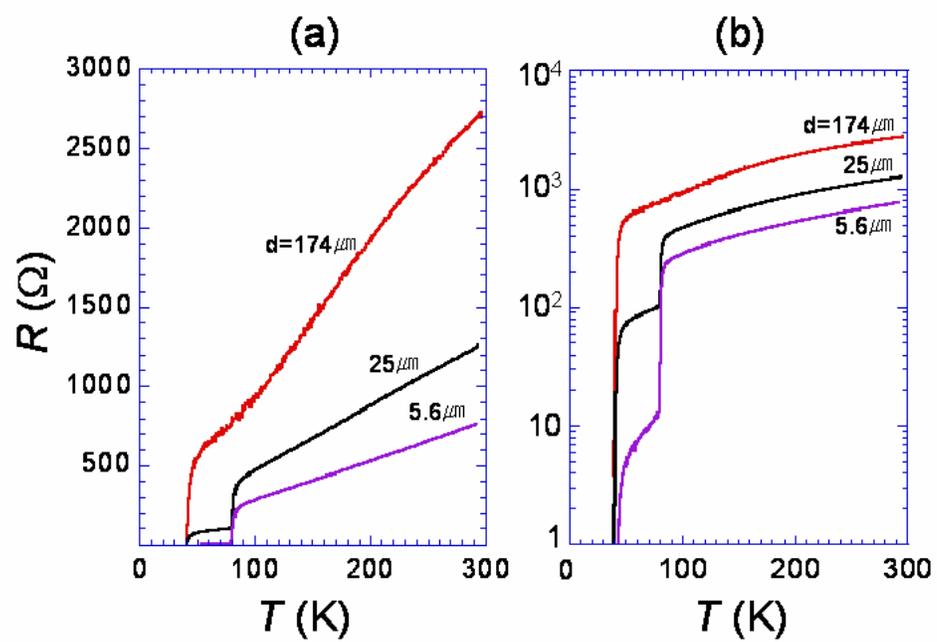